\documentclass[amsmath,amssymb,aps,prl,twocolumn,showpacs,superscriptaddress]{revtex4}

\usepackage{mathptmx}

\newcommand{\replace}[3]{#3}

\def \bx {{\bf x}}
\def \bp {{\bf p}}
\def \bn {{\bf n}}
\def \bm {{\overline{m}}}

\begin{document}
\title{
Relativistic Closed-Form Hamiltonian for Many-Body Gravitating Systems\\
in the Post-Minkowskian Approximation
}
\author{Tom\'a\v s Ledvinka}
\affiliation{Institute of Theoretical Physics, Faculty of Mathematics and Physics, Charles University, Prague, Czech Republic}
\affiliation{Theoretisch-Physikalisches Institut, Friedrich-Schiller-Universit\"at, Max-Wien-Platz\ 1, 07743 Jena, Germany}
\author{Gerhard Sch\"afer}
\affiliation{Theoretisch-Physikalisches Institut, Friedrich-Schiller-Universit\"at, Max-Wien-Platz\ 1, 07743 Jena, Germany}
\affiliation{Institute of Theoretical Physics, Faculty of Mathematics and Physics, Charles University, Prague, Czech Republic}
\author{Ji\v r\'\i\ Bi\v c\'ak}
\affiliation{Institute of Theoretical Physics, Faculty of Mathematics and Physics, Charles University, Prague, Czech Republic}
\affiliation{Theoretisch-Physikalisches Institut, Friedrich-Schiller-Universit\"at, Max-Wien-Platz\ 1, 07743 Jena, Germany}
\received{27 January 2008}
\pacs{04.25.-g, 04.25.Nx}
\begin{abstract}
The Hamiltonian for a system of relativistic bodies interacting by their gravitational field is found in~the
post-Minkowskian approximation, including all terms linear in the gravitational constant. It is given in a
surprisingly simple closed form as a function of canonical variables describing the bodies only. The field
is~eliminated by solving inhomogeneous wave equations, applying transverse-traceless projections, and
using the Routh functional. By including all special relativistic effects our Hamiltonian extends the results
described in classical textbooks of theoretical physics. As an application, the scattering of relativistic
objects is considered.
\end{abstract}
\maketitle

{\it Introduction.---}The problem of motion of bodies under
gravitational interaction has been the central issue in
general relativity from its birth. As early as 1916, Einstein,
Droste, Lorentz, and de Sitter started to develop post-
Newtonian (PN) approximation methods. These are based
on the weak-field limit in which the metric is close to the
Minkowski metric  and the assumption that the typical
velocity $v$ in a system divided by the speed of light $c$ is
very small, $v/c\sim \epsilon$. The deviation from the flat metric can
be characterized by the Newtonian potential $\Phi$; so for a
binary system, for example, $\Phi/c^2 \sim v^2/c^2\sim \epsilon^2$. In an
appropriate limit (as $\epsilon\rightarrow 0$), the PN approximation yields
Newton's equations. In 1916 Einstein also worked out ``the
linearized approximation'' to general relativity
in which the flat-space wave equations for the deviations $h_{\mu\nu}$ from
Minkowski metric with an energy-momentum tensor $T_{\mu\nu}$
of matter as the source are considered. This work marks
the beginning of the post-Minkowskian (PM) approximation methods: the weakness
of the gravitational field is assumed -- the deviations of the metric
from flat are small, but no assumption about slowness of motion is made.
In an appropriate limit, when a suitable parameter, usually identified with
Newton's gravitational constant $G$, is approaching zero,
the PM approximation yields equations of special relativity.
The ``historical'' period culminated in 1938 when
Einstein, Infeld, and Hoffmann, by formal expansions in $c^{-1}$,
investigated the $n$-th iterated field equations of the PN scheme.

Since the 1950s numerous investigations of both PN and
PM approximations appeared; see, for example,
refined comprehensive reviews \cite{Dam2x,Tho1x}. Because of the
evidence that the orbit of the binary pulsar PSR 1913+16
decays as a consequence of the emission of gravitational
waves, the most promising candidates for the detectors
such as LIGO, VIRGO, and GEO600 became binary
neutron stars or black holes. This led to new studies of higher-order
equations of motion. Extensive reviews summarizing
these developments were prepared recently \cite{Bl,FI}.

The Hamiltonian methods have been widely used in
general relativity in such problems as perturbation of black
holes, dynamics of anisotropic cosmological models, etc.
In the problem of motion of gravitating systems the
Hamiltonian yields equations of motion, expressions for
the mechanical energy and angular momentum; in the case
of binaries, the result for the last stable circular orbit
follows from $\partial^2 H/\partial r^2=0$. The Hamiltonian approach
to the PN approximation was initiated by Kimura and
Toiya \cite{KimuraToiya} and developed recently in the work of Sch\"afer,
Jaranowski and Damour (see \cite{FI} for references). The
canonical formalism of Arnowitt, Deser, and Misner
(ADM) \cite{ADM} is commonly used.

Before we enter into details we wish to indicate why our
final result--the closed form Hamiltonian (\ref{H1PM}) including
all terms linear in $G$---is  of importance in the theory of
gravity and can be relevant in other branches of theoretical
physics and astrophysics: (i) Since PM approximation can
be restricted to slow motions, the Hamiltonian describes
PN approximations to {\it any} order in $1/c$ when terms linear
in $G$ are considered.  When considering only two particles
and terms up to the second order ($\lesssim v^2/c^2$) in center-of-mass
system we precisely recover the Hamiltonian given
by Landau and Lifshitz \cite{LL}, except for the last term which
is $\sim G^2$. Moreover, we checked that, after a suitable canonical
transformation, it yields precisely all the terms
linear in $G$ in 3PN approximation of \cite{DJS00}; and we
calculated corresponding terms in 4PN order which have
not been given thus far. These will be published elsewhere.
(ii) An electromagnetic counterpart of our Hamiltonian
was derived by completely different methods by
Kennedy \cite{kennedy}.
The first terms in the expansion in $c^{-2}$
of the Kennedy Hamiltonian correspond to the well-known
Darwin Lagrangian (see e.g. \cite{LL,Jackson}) but Kennedy's
treatment goes beyond Darwin's approximation. Our procedure
shows how the Darwin Hamiltonian or Lagrangian can be
generalized in both gravitational and electromagnetic cases.
\replace{13}{ 
        The Darwin Lagrangian had most important
        applications in a quantum analysis of relativistic corrections
        in 2-electron atoms, but it also has uses in the purely
        classical domains such as a tool for electromagnetic
        plasma simulation, for statistical mechanics of many-body
        systems, etc. (see \cite{Jackson} for references). The extension
        for two point charges considered up to order $c^{-4}$ (included)
        has been given by the Golubenkov-Smorodinsky
        Lagrangian or Hamiltonian, which is described in \cite{LL} and
        discussed in \cite{DamourSchaefer}. Our Hamiltonian will give analogous
        results for gravity.
}{ 
        The Darwin Lagrangian or Hamiltonian have many applications in both
        classical and quantum domains (see, e.g., \cite{Jackson}).
        The extension for two point charges considered up to order $c^{-4}$ (included)
        has been given by the Golubenkov-Smorodinsky Lagrangian,
        which is described in \cite{LL} and discussed in \cite{DamourSchaefer}.
}
(iii) As our Hamiltonian can describe
particles with ultrarelativistic velocities or with zero rest
mass, it is well suited for treating gravitational scattering of
such objects. These are closely related to gravitational
scattering of shock waves. At high energies in quantum
scattering processes, the interaction is well approximated
by the classical collision at the speed of light, as first shown
by the influential work by 't~Hooft
(see \cite{DEath} for detailed description and methods).
(iv) Because of its simplicity,
our Hamiltonian yields a convenient starting point in
investigations of gravitational lensing by extended gravitationally
interacting objects since it provides a unified
description of both gravitating bodies and deflected
photons or neutrinos. (v) Our method of deriving the
Hamiltonian by using the Routhian and the induced
canonical transformation, etc. should be of interest in other
branches of theoretical physics.

The Hamiltonian approach to the  PM approximation
was first undertaken in 1986 \cite{S86}. Following the ADM
canonical formalism, the gravitational field is described
by $h_{ij}^{TT}$, the transverse-traceless part of $h_{ij}=g_{ij}-\delta_{ij}$
($h^{TT}_{ii}=0$, $h^{TT}_{ij,j}=0$, $i,j=1,2,3$), and by conjugate
momenta ${\bar \pi}^{ij\,TT}$. The system of bodies located at $\bx_a$, $a=1, ... N$, with rest masses $m_a$
and momenta $\bp_a$, has the energy and linear momentum densities
${\gamma}^\frac{1}{2} T^{\mu\nu}n_\mu n_\nu =\sum_a \left(g^{ij} p_{ai} p_{aj}+m_a^2\right)^\frac{1}{2} \delta(\bx-\bx_a)$,
$-{\gamma}^\frac{1}{2}T^{\;\mu}_i n_\mu =\sum_a p_{ai}  \delta(\bx-\bx_a)$,
where $\gamma=\det(g_{ij})$, $g^{ij}$ is inverse to $g_{ij}$,
$n^{\nu}$ is a unit timelike normal to hypersurface $x^0=\rm const$, and
$T^{\mu\nu}$ is the energy-momentum tensor of the matter system.
Hereafter, we call its constituents the ``particles'', but they
may well represent neutron stars or black holes.
This is substantiated by ``general relativity's adherence to the strong equivalence principle'': black holes and other bodies obey the same laws of motion as test bodies; see, e.g., \cite{DEath}.
\replace{14}{ 
        (Also, the analysis of the initial-value solutions for black holes shows that as in electromagnetism, where image charges are described by delta functions, black holes in full general relativity can be represented by ``image masses'' with delta functions support \cite{JS}.)
}{} 

It is convenient to choose four coordinate conditions $\bar \pi^{ii}=0$
and $g_{ij} = \left(1+\frac{1}{8}\phi \right)^4 \delta_{ij}+h_{ij}^{TT}$.
The standard ADM Hamiltonian (cf. \cite{ADM}),
$H = (1/16 \pi G) \oint dS_i (g_{ij,i}-g_{jj,i})$, then becomes, using the Gauss theorem,
$H=-(1/16 \pi G) \int d^3x \Delta \phi$. The integrand $\Delta \phi=\partial_i\partial_i \phi$ can be expressed in terms of $\bx_a$, $\bp_a$, $h^{TT}_{ij}$ and ${\bar \pi}^{ij\,TT}$
from the constraint equations. By expansions in powers of $G$
and after adopting suitable regularization procedures
of integrals involved (see the Appendix in \cite{S86}), one can determine the Hamiltonian.
The Hamiltonian worked out in \cite{S86} includes terms $\sim G^2$.
\replace{15}{ 
        ; the equations of motion following from the Hamiltonian are then correct also up to $G^2$.
}{} 

Here we start from the same Hamiltonian neglecting, however, terms $\sim G^2$.
Still, the  form of the Hamiltonian is quite complicated [see Eq. (\ref{HlinGS})]:
rather than just quantities associated with particles it involves field variables,
non-local TT projections, and integrals.
The main purpose of this Letter is to show that due to somewhat magical simplifications the fields
$h^{TT}_{ij}$, assuming they are generated just by particles,
can be expressed entirely in terms of particles' variables, and after
two canonical transformations, the Hamiltonian
can be cast into the closed form (\ref{H1PM}).
It is local in particles' canonical variables,
involving no field variables.
It is exact up to terms linear in $G$, and within the same accuracy it yields equations of motion for particles
moving with possibly ultrarelativistic speeds, including those with zero rest mass.
\replace{16}{
        Since PM approximation can be restricted to slow motions, the Hamiltonian
        describes PN approximations to {\it any} order in $1/c$ when terms linear in $G$ are considered.
}{}

{\it 1PM Hamiltonian.---}
The Hamiltonian describing $N$ particles ($\bx_a$, $\bp_a$) and
their gravitation field ($h_{ij}^{TT}$, $\pi^{ij\,TT}$)
accurate up to the terms linear in $G$ reads
\begin{align}
\label{HlinGS}
 H_{\rm lin}=&
\sum_a \bm_a - \frac{1}{2}G\sum_{a,b\ne a} \frac{\bm_a \bm_b }{ r_{ab} }
\left( 1+ \frac{p_a^2}{\bm_a^2}+\frac{p_b^2}{\bm_b^2}\right)
\\\nonumber
&+ \frac{1}{4}G\sum_{a,b\ne a} \frac{1}{r_{ab}}\left( 7\, \bp_a.\bp_b + (\bp_a.\bn_{ab})(\bp_b.\bn_{ab})   \right)
\\
\nonumber
&-\frac{1}{2}\sum_a \frac{p_{ai}p_{aj}}{\bm_a}\,h_{ij}^{TT}(\bx=\bx_a)
\\\nonumber
&s+\frac{1}{16\pi G}\int d^3x~ \left( \frac{1}{4} h_{ij,k}^{TT}\, h_{ij,k}^{TT} +\pi^{ij\,TT} \pi^{ij\,TT}\right)~,
\end{align}
where $\bm_a=\left( m_a^2+\bp^2_a \right)^\frac{1}{2}$,
$\bn_{ab} r_{ab} = \bx_a-\bx_b$, $|\bn_{ab}|=1$.
[This is Eq. (18) in \cite{S86}. However,
although we also put $c=1$,
we keep explicitly $G$ and do not put $16\pi G =1$;
the ``geometric'' momenta $\pi^{ij\,TT}$ are thus different from
the true ${\bar\pi}^{ij\,TT} = (16\pi G)^{-1} \pi^{ij\,TT}$. Notice that
the regularization is also needed in the terms containing $h_{ij}^{TT}$, $\pi^{ij\,TT}$,
and $h_{ij}^{TT}(\bx=\bx_a)$.]
The equations of motion for particles are standard Hamilton equations.
The Hamilton equations for the field are
\begin{equation}
 \dot {\bar \pi}^{ij\,TT}~=~-~\delta_{kl}^{TT\,ij} \frac{\delta H}{\delta h_{kl}^{TT}}
~,~~~
 \dot h_{ij}^{TT}~=~~\delta_{ij}^{TT\,kl} \frac{\delta H}{\delta {\bar \pi}^{kl\,TT}}~;
\end{equation}
here the variational derivatives and the TT-projection operator
$\delta_{kl}^{TT\,ij} = \frac{1}{2}\left( \Delta_{ik}\Delta_{jl}+\Delta_{il}\Delta_{jk}-\Delta_{ij}\Delta_{kl}\right){\Delta^{-2}}$,
 $\Delta_{ij} = \delta_{ij}\Delta - \partial_i\,\partial_j$,  appear.
These equations imply the equations for gravitational field
in the first PM approximation to be the following
wave equations:
\begin{equation}
\label{we4h}
\square h_{ij}^{TT} ~=~ - 16\pi G~ \delta_{ij}^{TT\,kl}
\sum_a \frac{p_{ak}p_{al}}{\bm_a }\delta^{(3)}( \bx-\bx_a)~.
\end{equation}
As is usual in the system of particles and field, the field is source of particles'
accelerations and particles are sources of the field.
Once $O(G^2)$ terms can be neglected, however, this coupling simplifies.
Since both the field and the accelerations $\dot \bp_a$
are proportional to $G$, the changes of the field due to the accelerations of particles are
of the order $O(G^2)$.
Therefore, we can assume the field to be generated by unaccelerated
particles; it is thus given only by their instantaneous positions and velocities.
If such a field is used in equations of motion of particles,
the gravitational interaction becomes an ``action at a distance''.

Given the linearity of the field equations (\ref{we4h}),
we can write $h^{TT}_{ij}(\bx) = \sum_b h^{TT}_{ij}(\bx;\bx_b,\bp_b,\dot \bx_b)$,
where $h_{ij}^{TT}(\bx ; \bx_b,\bp_b,\dot \bx_b)$ is the contribution
to the field at point $\bx$ generated by the particle
moving at $\bx_b$ with velocity $\dot \bx_b$.
Because the d'Alembertian operator commutes with the TT-projection operator,
the field from each particle involves the solution of the scalar wave equation with
point source -- the boosted static spherical field $\sim 1/r$.
Denoting $\bx-\bx_a=\bn_a |\bx-\bx_a|$ and $\cos \theta_a={\bn_a.\dot \bx_a /|\dot \bx_a|}$,
the solution of (\ref{we4h}) can thus be written as
\begin{equation}
\label{LiWi4h}
h_{ij}^{TT}(\bx) =
\delta_{ij}^{TT\,kl} \sum_b
\frac{4G}{\bm_b}
\frac{1}{|\bx-\bx_b|}
\frac{p_{bk}p_{bl}}{\sqrt {1-{\dot \bx_b}^2\sin^2 \theta_b}}~.
\end{equation}

The action of $\delta^{TT}_{ijkl}$ in equation (\ref{LiWi4h})
consists of two steps: first, one has to solve the Poisson equation twice
and then evaluate a number of partial derivatives.
The first step is feasible due to the form of the boosted spherical potential.
It can be shown to satisfy the relation (here $v=|\dot \bx_a|$)
\begin{align}
\Delta^{2} \left (
{ |\bx-\bx_a|}
{ \sqrt {1-v^2\sin^2 \theta_a}} \right )^3
=&\\\nonumber
3 (1-v^2)^2 \left[
8+7 v \frac{d}{dv}+v^2 \frac{d^2}{dv^2}
\right]&
\left(
\frac{1}{|\bx-\bx_a|}\frac{1}{\sqrt {1-v^2\sin^2 \theta_a}}
\right) .
\end{align}
Hence, instead of working out $\Delta^{-2}$,
we can evaluate $\left[8+7 v\; {d/dv}+v^2\; {d^2/dv^2}\right]^{-1}$,
i.e., the elliptic partial differential equation of the fourth order can be simplified
into an inhomogeneous linear second order ordinary differential equation.
It has a unique solution which is regular at $v=0$.
Even though this solution
is quite complicated and contains logarithmic terms such as $\ln\, \cos \theta_a$,
they cancel out when partial derivatives are combined.
After somewhat lengthy calculations (the details of which will be given elsewhere),
we find the field of a moving source
\begin{widetext}
\begin{align}
\nonumber
h_{ij}^{TT}(\bx;\bx_b,\bp_b,\dot \bx_b) ~=~ &
\frac{G}{|\bx-\bx_b|} \frac{1}{\bm_b}\frac{1}{y(1+y)^2}
\Big\{
\left[y\bp_b^2-(\bn_b.\bp_b)^2(3y+2)\right]\delta_{ij}
+2\left[
1- \dot \bx_b^2(1 -2\cos^2 \theta_b)\right]{p_{bi}p_{bj}}
\\\nonumber&
+\left[
\left( 2+y\right)(\bn_b.\bp_b)^2
\!-\!\left( 2+{3}y -2\dot \bx_b^2\right)\bp_b^2
\right]{n_{bi}n_{bj}}
+2(\bn_b.\bp_b) \left(1-\dot \bx_b^2+2y\right) \left(n_{bi}p_{bj}+p_{bi}n_{bj}\right)
\Big\}
\\
&+O(\bm_b \dot \bx_b-\bp_b)G\!+\!O(G^2)~;
\label{unprojected_h}
\end{align}
\end{widetext}
here $y = y_b \equiv\sqrt {1-{\dot \bx_b}^2\sin^2 \theta_b}$.
As indicated by the symbol $O(\bm_b \dot \bx_b-\bp_b)$, the last expression gets simplified by using
$\dot \bx_b=\bp_b/\bm_b$
and we anticipate that later
$O(\bm_b \dot \bx_b-\bp_b)G$ will turn into terms $\sim O(G^2)$.

In order to later suppress field degrees of freedom,
we shall turn to
the Routh functional (see, e.g., \cite{DJS00})
\begin{equation}
R( \bx_a,\bp_a, h_{ij}^{TT}, \dot h_{ij}^{TT} ) =
H - \frac{1}{16\pi G}\int d^3x~  \pi^{TT\,ij}\, \dot h_{ij}^{TT}
\end{equation}
which is ``the Hamiltonian for the particles but the Lagrangian for the field.''
While the functional derivatives of the Hamiltonian yield the time derivatives of the canonically conjugated field, the functional derivatives of Routhian vanish if the field equations (\ref{we4h}) hold. Their solution (\ref{LiWi4h})
is non-radiative and
can thus be substituted into the Routh functional without changing the Hamilton equations for the particles.
Hereafter the symbol $h^{TT}_{ij}$
is a shortcut for the solution of (\ref{we4h}) depending on coordinates, momenta and velocities
of the particles.
\replace{24}{
        So the reduced Hamiltonian $H(\bx_a,\bp_a, \dot\bx_a)$ is obtained.
        Its field part,
}{
        So the reduced Routhian (7) which becomes Hamiltonian $H(\bx_a,\bp_a, \dot\bx_a)$ is obtained.
        The field part of the Routhian
}
\begin{equation}
R_f ~=~ \frac{1}{16\pi G} \int d^3x~   \frac{1}{4}\left( h_{ij,k}^{TT}\, h_{ij,k}^{TT} - \dot h_{ij}^{TT}\dot h_{ij}^{TT} \right)~,
\end{equation}
still needs to be transformed into an explicit function of the particles' variables.
Using Gauss's law in the first term and integrating by parts the second term, we arrive at
\begin{align}
\label{HfGauss}
R_f &~=~ -\frac{1}{16\pi G}\int d^3x~  \frac{1}{4} h_{ij}^{TT}\left( \Delta  h_{ij}^{TT}
      - \frac{\partial^2}{\partial t^2} h_{ij}^{TT} \right)~~~~~~~~~~
\\\nonumber
&+ \frac{1}{64\pi G} \oint dS_k  (h_{ij}^{TT} h_{ij,k}^{TT})
 - \frac{1}{64\pi G} \frac{d}{dt} \int d^3 x~(h_{ij}^{TT} \dot h_{ij}^{TT})~.
\end{align}
The field equations (\ref{we4h}) imply that the first integral
(in which the self-interaction term is discarded) directly combines
with the ``interaction'' term containing
$\sum\, \bm_a^{-1}\, p_{ai}\, p_{aj} \,h^{TT}_{ij}(\bx_a)$, so only its coefficient is changed.
\replace{25}{
        Remaining terms in $H_f$, do not modify the dynamics of the system
        since in our approximation of unaccelerated -- and non-radiating --
        particles the surface integral vanishes at large $|\bx|$.
}{
        Remaining terms do not modify the dynamics of the system in our approximation.
}
The Hamiltonian thus takes the form
\begin{widetext}
\begin{align}
\nonumber
H_{\rm lin}(\bx_c,\bp_c,\dot \bx_c)=&
\sum_a \bm_a - \frac{1}{2}G\sum_{a,b\ne a} \frac{\bm_a \bm_b }{ r_{ab}}
s\left( 1+ \frac{p_a^2}{\bm_a^2}+\frac{p_b^2}{\bm_b^2}\right)
+ \frac{1}{4}G\sum_{a,b\ne a} \frac{1}{r_{ab}}\left( 7\, \bp_a.\bp_b + (\bp_a.\bn_{ab})(\bp_b.\bn_{ab})   \right)
\\ \label{Hlin}
&-\frac{1}{4}\sum_a \frac{p_{ai}p_{aj} }{\bm_a}\,h_{ij}^{TT}(\bx=\bx_a;\bx_b,\bp_b,\dot \bx_b)~.
\end{align}
Dropping out the total time derivatives in (\ref{HfGauss}) means
a canonical transformation, but the new canonical coordinates will not be denoted by primes.
In fact, another change of coordinates has to follow since the Hamiltonian (\ref{Hlin}) is a function of $\dot \bx_a$.
We define new momenta by putting
$p'_{ai}=p_{ai}-{\partial H}/{\partial \dot x_{ai}}$,
and then eliminate $\dot \bx_a$ by introducing new Hamiltonian
$ H'(\bx_a,\bp'_a)=H\!\left(\bx_a,\bp_a(\bp'_a),\dot \bx_a(\bp'_a)\right) -\!
\sum_b \dot x_{bi}(\bp'_a) \, {\partial H}/{\partial \dot x_{bi}}$.
Since ${\partial H/ \partial \dot x_a}\sim G$, the only change in the Hamiltonian
which is  linear in $G$ comes into the kinetic term $\sum \bm_a$ from the last change of momenta.
This change is exactly cancelled by the sum
$\sum_b \dot x_{bi}(\bp'_a) \, {\partial H}/{\partial \dot x_{bi}}$ in $ H'(\bx_a,\bp'_a)$.
We now make simple substitutions
$\dot x_{ai}=\frac{\partial \bm_a }{ \partial p_{ai}}= \frac{p_{ai}}{ \bm_a}
$,
$\cos \theta_a = \frac{\bn_a.\bp_a }{ |\bp_a|}~,
y_a^2\equiv1 - \dot \bx_a^2 \sin^2 \theta_a =
1+{\bm_a}^{-2}\left[ \left (\bn_a.\bp_a\right)^2-\bp_a^2 \right]~$
and, again, omit primes.
At this moment we can substitute for $h_{ij}^{TT}$ in (\ref{Hlin}) the solution (\ref{unprojected_h}),
in which the above substitutions turn the term $O(\bm_b \dot \bx_b-\bp_b)G$ into term $\sim O(G^2)$.
In this way, using the shortcut $y_{ba} = \bm_b^{-1} [ m_b^2+ \left (\bn_{ba}.\bp_b\right)^2]^\frac{1}{2}$,
we finally arrive at the Hamiltonian in the first post-Minkowskian approximation:
\begin{align}
\label{H1PM}
H_{\rm lin}~=&~
\sum_a \bm_a - \frac{1}{2}G\sum_{a,b\ne a} \frac{\bm_a \bm_b }{ r_{ab}}
\left( 1+ \frac{p_a^2}{ \bm_a^2}+\frac{p_b^2}{\bm_b^2}\right)
+ \frac{1}{4}G\sum_{a,b\ne a} \frac{1}{r_{ab}}\left( 7\, \bp_a.\bp_b + (\bp_a.\bn_{ab})(\bp_b.\bn_{ab})   \right)
\\ \nonumber
&-\frac{1}{4}
G\sum_{a,b\ne a} \frac{1}{r_{ab}}
\frac{(\bm_a \bm_b)^{-1}}{ (y_{ba}+1)^2 y_{ba}}
\Bigg[
2\Big(2
(\bp_a. \bp_b)^2 (\bp_b.\bn_{ba})^2
-2 (\bp_a.\bn_{ba}) (\bp_b.\bn_{ba}) (\bp_a.\bp_b) \bp_b^2
+(\bp_a.\bn_{ba})^2 \bp_b^4
-(\bp_a. \bp_b)^2 \bp_b^2
\Big ) \frac{1}{\bm_b^2}
\\ \nonumber
&+2 \left[-\bp_a^2 (\bp_b.\bn_{ba})^2 + (\bp_a.\bn_{ba})^2 (\bp_b.\bn_{ba})^2 +
2 (\bp_a.\bn_{ba}) (\bp_b.\bn_{ba}) (\bp_a.\bp_b) +
(\bp_a.\bp_b)^2 - (\bp_a.\bn_{ba})^2 \bp_b^2\right]
\\ \nonumber
&+
{\left[-3  \bp_a^2 (\bp_b.\bn_{ba})^2 +(\bp_a.\bn_{ba})^2 (\bp_b.\bn_{ba})^2
+8 (\bp_a.\bn_{ba}) (\bp_b.\bn_{ba}) (\bp_a.\bp_b)
+ \bp_a^2 \bp_b^2 - 3 (\bp_a.\bn_{ba})^2 \bp_b^2 \right]y_{ba}
}
\Bigg]~.
\end{align}
\end{widetext}
This is our main result. The Hamiltonian for a many-particle gravitating system
in post-Minkowskian approximation, i.e., including all terms linear in $G$,
was derived in the closed form entirely in terms of the variables of the particles.
Putting $G=0$ it becomes standard Hamiltonian of $N$ noninteracting particles
in special relativity.

Let us yet note that in
\cite{Fr} the post-Minkowskian action for a helically symmetric binary solution was constructed;
however, it turns out that due to the restriction to helical worldlines ambiguities may arise.
We will return to this issue elsewhere.
It should be also useful to derive the Hamiltonian (\ref{H1PM}) within the
effective field theory approach to gravity \cite{GR}.

{\it Scattering.---} As an application of the Hamiltonian obtained we calculate
gravitational scattering of two possibly ultrarelativistic or zero-rest-mass particles.
When only terms linear in $G$ are considered, the transferred momentum can be computed as
$
\Delta \bp_1 = \int_{-\infty}^{+\infty} {\dot \bp_1} ~dt~,
$
integrating along the straight line trajectories of noninteracting particles
[${\dot \bp_1}$ is determined from the Hamilton equations using (\ref{H1PM})].
If perpendicular separation $\bf b$ of trajectories ($|{\bf b}|$ is the impact factor)
in center-of-mass system ($\bp_1=-\bp_2\equiv\bp$) is used, $\bp.{\bf b}=0$,
we find, after evaluating a few simple integrals, that the exchanged momentum in the
system is given by
\begin{align}
\label{delta_p}
\Delta \bp &= -2\frac{{\bf b}}{{\bf b}^2} \frac{G}{|\bp|}
\frac{\bm_1^2 \bm_2^2}{\bm_1 +\bm_2 }
\\\nonumber
\times
&\left[
1+\left(\frac{1}{\bm_1^2}+\frac{1}{\bm_2^2}+\frac{4}{\bm_1\bm_2} \right)\bp^2
+\frac{\bp^4}{\bm_1^2 \bm_2^2 }
\right]~.
\end{align}
The quartic term is all that remains from the field part $h^{TT}_{ij}$.
It is not difficult to show that (\ref{delta_p}) agrees with the result \cite{W85}
obtained by a very different method.

\vskip 2mm
The authors benefitted from the exchange program between Charles University, Prague,
and Friedrich Schiller University, Jena.
T.L. and J.B. acknowledge the partial support from SFB/TR7 in Jena,
from the Grant GA\v CR 202/06/0041 of the Czech Republic, and of Grants No LC 06014 and the MSM 0021620860
of Ministry of Education. J.B. is also grateful for the support of the Alexander von Humboldt Foundation.


\begin{thebibliography}{11}
\expandafter\ifx\csname natexlab\endcsname\relax\def\natexlab#1{#1}\fi
\expandafter\ifx\csname bibnamefont\endcsname\relax
  \def\bibnamefont#1{#1}\fi
\expandafter\ifx\csname bibfnamefont\endcsname\relax
  \def\bibfnamefont#1{#1}\fi
\expandafter\ifx\csname citenamefont\endcsname\relax
  \def\citenamefont#1{#1}\fi
\expandafter\ifx\csname url\endcsname\relax
  \def\url#1{\texttt{#1}}\fi
\expandafter\ifx\csname urlprefix\endcsname\relax\def\urlprefix{URL }\fi
\providecommand{\bibinfo}[2]{#2}
\providecommand{\eprint}[2][]{\url{#2}}

\bibitem[{\citenamefont{Damour}(1983)}]{Dam2x}
\bibinfo{author}{\bibfnamefont{T.}~\bibnamefont{Damour}}, in
  \emph{\bibinfo{booktitle}{Gravitational Radiation}}, edited by
  \bibinfo{editor}{\bibfnamefont{N.}~\bibnamefont{Deruelle}} \bibnamefont{and}
  \bibinfo{editor}{\bibfnamefont{T.}~\bibnamefont{Piran}}
  (\bibinfo{publisher}{North-Holland}, \bibinfo{address}{Amsterdam},
  \bibinfo{year}{1983}), pp. \bibinfo{pages}{59--144}; \bibinfo{note}{in
  {\it Three Hundred Years of Gravitation}, edited by S. Hawking and W. Israel,
  (Cambridge University Press, Cambridge, England, 1987), pp. 128-198}.

\bibitem[{\citenamefont{Thorne}(1987)}]{Tho1x}
\bibinfo{author}{\bibfnamefont{K.}~\bibnamefont{Thorne}}, in
  \emph{\bibinfo{booktitle}{Three Hundred Years of Gravitation}} (Ref. [1]), 
  pp. \bibinfo{pages}{330--458}.

\bibitem[{\citenamefont{Blanchet}(2002)}]{Bl}
\bibinfo{author}{\bibfnamefont{L.}~\bibnamefont{Blanchet}},
  \bibinfo{journal}{Living Reviews in Relativity} \textbf{\bibinfo{volume}{9}}, 4
  (\bibinfo{year}{2006})
  \url{http://www.livingreviews.org/lrr-2006-4}.

\bibitem[{\citenamefont{Futamase and Itoh}(2007)}]{FI}
\bibinfo{author}{\bibfnamefont{T.}~\bibnamefont{Futamase}} \bibnamefont{and}
  \bibinfo{author}{\bibfnamefont{Y.}~\bibnamefont{Itoh}},
  \bibinfo{journal}{Living Reviews in Relativity} \textbf{\bibinfo{volume}{10}}, 2
  (\bibinfo{year}{2007}),
  \url{http://www.livingreviews.org/lrr-2007-2}.

\bibitem[{\citenamefont{Kimura and Toiya}(1975)}]{KimuraToiya}
\bibinfo{author}{\bibfnamefont{T.} \bibnamefont{Kimura}} \bibnamefont{and}
\bibinfo{author}{\bibfnamefont{T.} \bibnamefont{Toiya}},
  \bibinfo{journal}{Prog. Theor. Phys.} \textbf{\bibinfo{volume}{48}},
  \bibinfo{pages}{316} (\bibinfo{year}{1972}).


\bibitem[{\citenamefont{Arnowitt et~al.}(1962)\citenamefont{Arnowitt, Deser,
  and Misner}}]{ADM}
\bibinfo{author}{\bibfnamefont{R.}~\bibnamefont{Arnowitt}},
  \bibinfo{author}{\bibfnamefont{S.}~\bibnamefont{Deser}}, \bibnamefont{and}
  \bibinfo{author}{\bibfnamefont{C.~W.} \bibnamefont{Misner}}, in
  \emph{\bibinfo{booktitle}{Gravitation: An Introduction to Current Research}},
  edited by \bibinfo{editor}{\bibfnamefont{L.}~\bibnamefont{Witten}}
  (\bibinfo{publisher}{Wiley}, \bibinfo{address}{New York},
  \bibinfo{year}{1962}), pp. \bibinfo{pages}{227--265}.

\bibitem[{\citenamefont{Landau and Lifshitz}(1975)}]{LL}
\bibinfo{author}{\bibfnamefont{L.~D.}~\bibnamefont{Landau}} \bibnamefont{and}
\bibinfo{author}{\bibfnamefont{E.~M.}~\bibnamefont{Lifshitz}},
  \emph{\bibinfo{booktitle}{The Classical Theory of Fields}}, 4th ed.
  (\bibinfo{publisher}{Pergamon},
  \bibinfo{address}{New York}, \bibinfo{year}{1975}).

\bibitem[{\citenamefont{Damour et~al.}(2000)\citenamefont{Damour, Jaranowski,
  and Sch{\"{a}}fer}}]{DJS00}
\bibinfo{author}{\bibfnamefont{T.}~\bibnamefont{Damour}},
  \bibinfo{author}{\bibfnamefont{P.}~\bibnamefont{Jaranowski}},
  \bibnamefont{and}
  \bibinfo{author}{\bibfnamefont{G.}~\bibnamefont{Sch{\"{a}}fer}},
  \bibinfo{journal}{Phys. Rev. D} \textbf{\bibinfo{volume}{62}},
  \bibinfo{pages}{021501(R)} (\bibinfo{year}{2000}); \bibinfo{note}{\,Phys.
  Rev. D {\bf 63} 029903(E) (2000)};
  \bibinfo{journal}{Phys. Lett. B} \textbf{\bibinfo{volume}{513}},
  \bibinfo{pages}{147} (\bibinfo{year}{2001}).

\bibitem[{\citenamefont{Kennedy}(1975)}]{kennedy}
\bibinfo{author}{\bibfnamefont{F.~J.} \bibnamefont{Kennedy}},
  \bibinfo{journal}{J. Math. Phys. (N.Y.)} \textbf{\bibinfo{volume}{16}},
  \bibinfo{pages}{1844} (\bibinfo{year}{1975}).

\bibitem[{\citenamefont{Jackson}(1975)}]{Jackson}
\bibinfo{author}{\bibfnamefont{J.~D.}~\bibnamefont{Jackson}},
  \emph{\bibinfo{booktitle}{ Classical Electrodynamics}}, 
  (\bibinfo{publisher}{John Wiley},
  \bibinfo{address}{New York}, \bibinfo{year}{1975}), 2nd ed.

\bibitem[{\citenamefont{Damour and Sch{\"{a}}fer}(1986)}]{DamourSchaefer}
\bibinfo{author}{\bibfnamefont{T.}~\bibnamefont{Damour}} \bibnamefont{and}
\bibinfo{author}{\bibfnamefont{G.}~\bibnamefont{Sch{\"{a}}fer}},
  \bibinfo{journal}{J. Math. Phys.  (N.Y.)} \textbf{\bibinfo{volume}{32}},
  \bibinfo{pages}{127} (\bibinfo{year}{1991}).


\bibitem[{\citenamefont{D'Eath}(1996)}]{DEath}
\bibinfo{author}{\bibfnamefont{P.~D.}~\bibnamefont{D'Eath}},
  \emph{\bibinfo{booktitle}{Black Holes: Gravitational Interaction}},
  (\bibinfo{publisher}{Clarendon Press},
  \bibinfo{address}{Oxford, England}, \bibinfo{year}{1996}).


\bibitem[{\citenamefont{Sch{\"{a}}fer}(1986)}]{S86}
\bibinfo{author}{\bibfnamefont{G.}~\bibnamefont{Sch{\"{a}}fer}},
  \bibinfo{journal}{Gen. Relativ. Gravit.} \textbf{\bibinfo{volume}{18}},
  \bibinfo{pages}{255} (\bibinfo{year}{1986}).


\bibitem[{\citenamefont{Friedman and Uryu}(2006)}]{Fr}
\bibinfo{author}{\bibfnamefont{J.~L.} \bibnamefont{Friedman}} \bibnamefont{and}
  \bibinfo{author}{\bibfnamefont{K.}~\bibnamefont{Uryu}},
  \bibinfo{journal}{Phys. Rev. D} \textbf{\bibinfo{volume}{73}},
  \bibinfo{pages}{104039} (\bibinfo{year}{2006}).

\bibitem[{\citenamefont{Friedman and Uryu}(2006)}]{GR}
\bibinfo{author}{\bibfnamefont{W.~D.} \bibnamefont{Goldberger}} \bibnamefont{and}
  \bibinfo{author}{\bibfnamefont{I.~Z.}~\bibnamefont{Rothstein}},
  \bibinfo{journal}{Phys. Rev. D} \textbf{\bibinfo{volume}{73}},
  \bibinfo{pages}{104029} (\bibinfo{year}{2006});
  \bibinfo{journal}{Gen. Relativ. Gravit.} \textbf{\bibinfo{volume}{38}},
  \bibinfo{pages}{1537} (\bibinfo{year}{2006}).


\bibitem[{\citenamefont{Westpfahl}(1985)}]{W85}
\bibinfo{author}{\bibfnamefont{K.}~\bibnamefont{Westpfahl}},
  \bibinfo{journal}{Fortschr. Physik} \textbf{\bibinfo{volume}{33}},
  \bibinfo{pages}{417} (\bibinfo{year}{1985}).

\end{thebibliography}
\end{document}